# Observation of high-order imaginary Poynting momentum optomechanics in structured light

Yuan Zhou[1,2]†, Xiaohao Xu[3]*†, Yanan Zhang[1,2], Manman Li[1], Shaohui Yan[1]*, Manuel Nieto-Vesperinas[4], Baojun Li[3], Cheng-Wei Qiu[5]* Baoli Yao[1,2]*

[1]*State Key Laboratory of Transient Optics and Photonics, Xi'an Institute of Optics and Precision Mechanics, Chinese Academy of Sciences, Xi'an 710119, China*
[2]*University of Chinese Academy of Sciences, Beijing 100049, China*
[3]*Institute of Nanophotonics, Jinan University, Guangzhou 511443, China*
[4]*Instituto de Ciencia de Materiales de Madrid, Consejo Superior de Investigaciones Científicas, Campus de Cantoblanco, Madrid 28049, Spain.*
[5]*Department of Electrical and Computer Engineering, National University of Singapore, Singapore 117583, Singapore*

*Corresponding authors: xuxhao@jnu.edu.cn, shaohuiyan@opt.ac.cn, chengwei.qiu@nus.edu.sg, yaobl@opt.ac.cn

†These authors contributed equally to this work.

**The imaginary Poynting momentum (IPM) of light has been captivated an unusual origin of optomechanical effects on dipolar magnetoelectric particles, but yet observed in experiments. Here, we report, for the very first time, a whole family of high-order IPM forces for not only magnetoelectric but also generic Mie particles, assisted with their excited higher multipoles within. Such optomechanical phenomena derive from a nonlinear contribution of the IPM to the optical force, and can be remarkable even when the incident IPM is small. We observe the high-order optomechanics in a structured light beam with vortex-like IPM streamlines, which allows the low-order dipolar contribution to be suppressed. Our results provide the first unambiguous evidence of the ponderomotive nature of the IPM, expand the classification of optical forces and open new possibilities for optical forces and micromanipulations.**

**Introduction**

The complex Poynting vector, $\mathbf{\Pi} = (\mathbf{E}^* \times \mathbf{H})/2$, is a fundamental characteristic quantity of Maxwell waves [1], and it plays a crucial role in light-matter interactions. For example, its real part, Re($\mathbf{\Pi}$), represents the density of time-averaged electromagnetic energy flux, and determines, up to a dimensional constant, an important dynamic property of light, namely, its time-averaged momentum. When light interacts with small objects, the incident momentum gives rise to the radiation pressure with implications spanning over the areas of laser cooling [2,3], optical micromanipulation [4] and optomechanical systems [5].

The physical meaning of Re($\mathbf{\Pi}$) is so straightforward that its imaginary counterpart, Im($\mathbf{\Pi}$), has sometimes been forgotten or even neglected in optical physics. The flow of the latter is usually interpreted as the reactive power [1], which used to be a topic only of interest to engineering researchers in RF-antenna design and in the electric power industry [6-8]. However, the imaginary Poynting vector has, in recent years, gained a growing interest in the burgeoning area of nanophotonics, and it is associated with the concept of imaginary Poynting momentum (IPM) [9-12]. This increased attention can be traced primarily to its influence in optical force theory [13], which shows that the IPM of the illumination can be coupled to a recoiling force via the interplay between electric and magnetic dipoles induced in the particles. Because the IPM force, in general, is linearly independent of the optical radiation pressure and intensity-gradient force, it offers a distinct degree of freedom for optically manipulating particles. Such a profound implication has been theoretically highlighted in different configurations, like optical tweezers [14], vector beams [12,15], evanescent [9,16] and two-wave interference fields [10,16,17]. Moreover, because the generation of this force is accompanied with asymmetric light scattering by the particle [13], it could be exploited for the design of advanced optical nanoantennas and sources [18-20].

Despite of these advances, there is still limited experimental evidence of the IPM force, which is often masked in the overall mechanical effect of light. We note that a recent work has made an effort to detect this force, by observing the deflection of a nanocantilever probe in the evanescent field [11]. However, the deflection could not directly signal the IPM force, due to the coexisting, stronger transverse radiation pressure [9,11], and potential torque effects caused by the anisotropy of the rod-like probe [21,22].

On the other hand, so far the knowledge of the IPM force is well established only for those particles that can be treated as magnetoelectric dipoles, but nothing is known of its role beyond the dipole approximation. In this context, we are naturally led to the following questions: (i) How to discriminate and thus observe the IPM force independently? (ii) Can the IPM be coupled to the optical force via multipolar effects? Then, can this force be observed in particles with no magnetic (or electric) response? Under which circumstances?

Regarding (i), one may seek for a field in which the IPM can be separated from all other field quantities that could have mechanical effects. We have previously shown a paraxial cylindrical vector (CV) beam, with IPM streamlines looping around the beam axis and orthogonal to both the optical momentum and intensity gradient [12]. Such a vortex structure therefore serves as a potential candidate to test the IPM force in the azimuthal direction. In this paper, we will deal with the IPM vortex of tightly focused (or nonparaxial) beams that can be applied to practical optical trapping and manipulation experiments. We also establish a useful multipolar model of optical force, which accounts for the relationship of the field quantities (including the IPM) with the force due to the multipolar interplay. This reveals a nonlinear field-to-force coupling, which is absent in the dipole approximation and leads to the high-order IPM force.

**Results**

**IPM vortex in tightly focused beams.** The geometry of the focusing problem is depicted in Fig. 1a, where a CV illumination with free space wavelength $\lambda$ is focused by a high numerical aperture (NA) objective lens into a non-dissipative medium, with permittivity $\varepsilon$ and permeability $\mu$. The input electric field can be written in the pupil plane polar coordinates $(\varrho, \varphi)$ as [23]

$$\mathbf{A}_0(\varrho,\varphi) = (\cos\alpha\, \mathbf{e}_\varrho + \sin\alpha\, \mathbf{e}_\varphi)\mathcal{A}_0(\vartheta) \qquad (1)$$

where $(\mathbf{e}_\varrho, \mathbf{e}_\varphi)$ are unit vectors along radial and azimuthal directions, and $\mathcal{A}_0(\vartheta)$ denotes the radial dependence of the field amplitude, with $\vartheta = \arcsin(\varrho/f)$ and $f$ referring to the focal length; the parameter $\alpha \in [-90°, 90°]$ characterizes the polarization state of the incident field: $\alpha = 0$ and $\pm 90°$ correspond to the typical radial and azimuthal polarizations, respectively. Throughout the paper, $\mathcal{A}_0(\vartheta)$ is taken to be real-valued as also employed in our experiments.

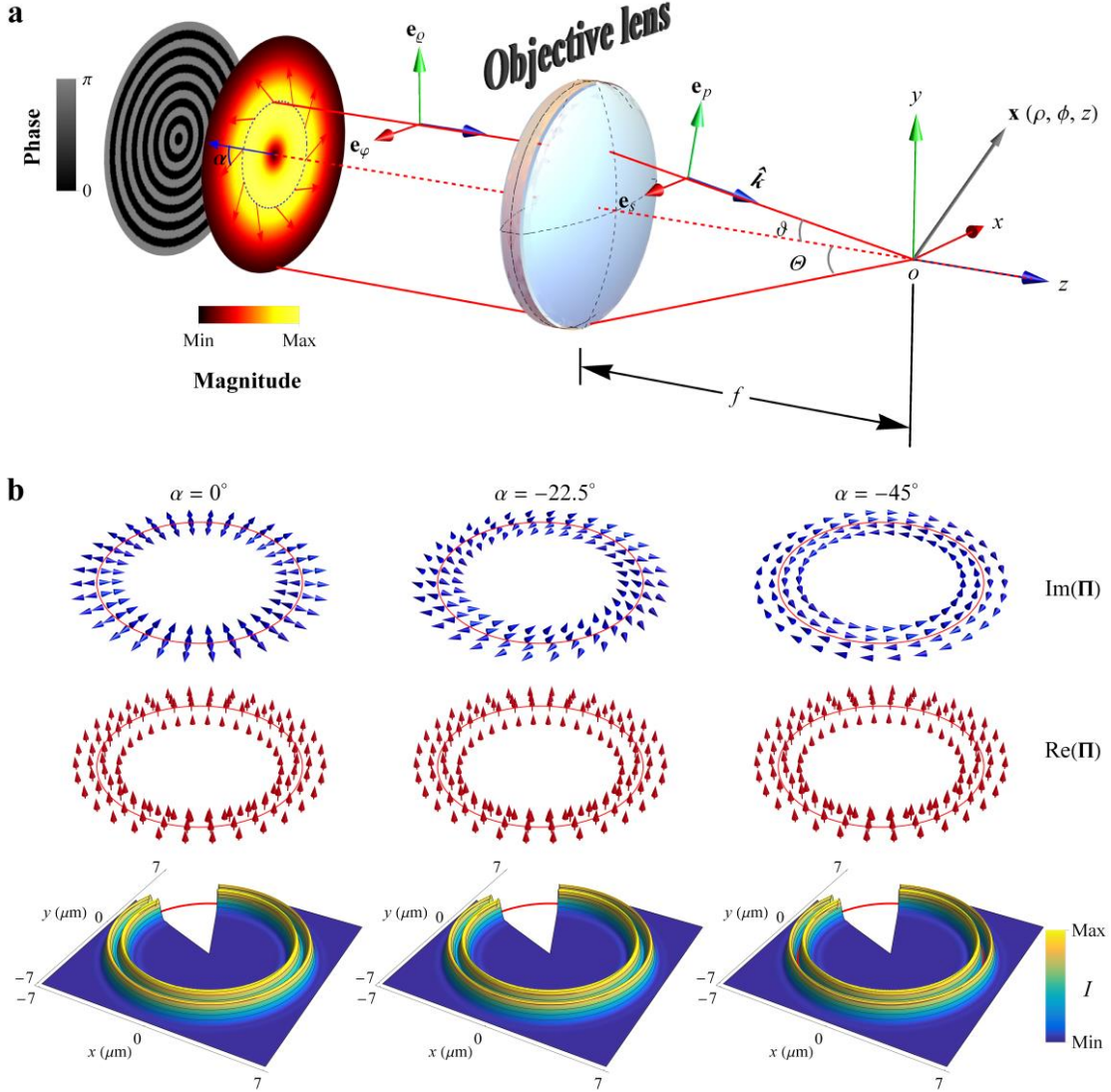

**Figure 1**. The IPM in nonparaxial CV beams. **a**, Geometry for analyzing the focusing property of CV beams. **b**, Calculated fields ($\lambda = 1.064$ μm) focused with NA = 1.26, for different polarization parameter $\alpha$. The refractive index of the ambient medium in image space is assumed to be $n_{\text{med}} = 1.33$ (water). Red circles with radius of 5.5 μm indicate the positions of local minima between intensity maxima.

Upon apodization, the field can be expressed by its $s$- and $p$-polarization components along $\mathbf{e}_s$ and $\mathbf{e}_p$, and employing the Richards-Wolf integral [24-26], one

may formulate the electric vector in the vicinity of the focus as (see Supplementary Note 1 for details)

$$\mathbf{E}(\mathbf{x}) = -i\frac{\cos\alpha}{k}\partial_z U(\rho,z)\mathbf{e}_\rho + U(\rho,z)\sin\alpha\,\mathbf{e}_\phi + i\frac{\cos\alpha}{k}\left(\partial_\rho + \frac{1}{\rho}\right)U(\rho,z)\mathbf{e}_z, \quad (2)$$

where $(\mathbf{e}_\rho, \mathbf{e}_\phi, \mathbf{e}_z)$ are the unit vectors for the cylindrical coordinates $\mathbf{x} = (\rho, \phi, z)$ in image space, and $k$ is the wavenumber in medium; $U$ is a function given by:

$$U(\rho,z) = -kf\int_0^\Theta \sqrt{\cos\vartheta}\,\mathcal{A}_0(\vartheta)J_1(k\rho\sin\vartheta)e^{ikz\cos\vartheta}\sin\vartheta d\vartheta, \quad (3)$$

with $J_1(\cdot)$ the Bessel function of the first kind of order one, and $\Theta = \arcsin(\mathrm{NA}/n_{\mathrm{med}})$ the maximal converging angle ($n_{\mathrm{med}}$ the refractive index of ambient medium).

Equation (2) holds exactly without any approximation [25]. With (2) and Maxwell's equations, it is shown that the azimuthal component $\Pi_\phi$ of the complex Poynting vector is purely imaginary over the whole focal volume (see Supplementary Note 1), and the imaginary Poynting vector at the focal plane takes the form

$$\mathrm{Im}(\mathbf{\Pi}) = \frac{1}{2\mu c}\left\{\frac{\cos 2\alpha}{k}\left(\frac{1}{\rho}|U|^2 + \frac{1}{2}\partial_\rho|U|^2\right)\mathbf{e}_\rho + \frac{\sin 2\alpha}{k^2}\mathrm{Im}\left[\frac{1}{\rho}\partial_\rho(\rho U^*)\partial_z U\right]\mathbf{e}_\phi\right\}, \quad (4)$$

where $c$ is the light speed in the medium, and we have made use of $\mathrm{Im}(U) = 0$, which holds for the field at the focal plane. *Equation (4) is a main theoretical result of this paper*, which describes the rotational dynamics of the IPM for a generic CV beam. Applying the paraxial approximation $\partial_z U \sim ikU$, one can produce from Eq. (4) the result for paraxial CV beams [12]. From the $\alpha$-dependent terms, it is deduced that the IPM will form a vortex structure at $\alpha = \pm 45°$, for which $\mathrm{Im}(\Pi_\phi)$ assumes the maximal value and $\mathrm{Im}(\Pi_\rho)$ is zero.

As an illustration, Fig. 1b shows the calculated focused fields with different polarization parameters. The amplitude function $\mathcal{A}_0(\vartheta)$ of the input field has a magnitude of standard CV beams, but modulated by a radially-varying phase profile (see insets in Fig. 1a, or Methods for details) to prevent the focused field from being localized near the beam axis. We see that the field intensity distribution, in either case, is characterized by a sharp dual ring-like profile. As the field is axially symmetrical, its intensity gradient should exist only in the radial direction. The longitudinal real Poynting vector, $\mathrm{Re}(\mathbf{\Pi})$, indicates that the field carries no net angular momentum. However, the IPM, $\mathrm{Im}(\mathbf{\Pi})$, has azimuthal component for $\alpha \neq 0$. Specially, for $\alpha = -45°$ this momentum circulates around the axis in a bidirectional manner, as shown in Fig. 1b (top right panel) where the IPM is clearly in opposite directions on the inner and outer sides of the annular field. Such a bi-chiral vortex structure is attributed to the spatial dependence of $\mathrm{Im}(\Pi_\phi)$, whose sign is changed by the radial position $\rho$ [see Eq. (4)].

**Multipolar effects on the IPM force.** An unambiguous observation of the IPM mechanical action requires comprehensive knowledge of the IPM force. In the dipole limit, this force can be written as [13]

$$\mathbf{F}_{\text{IPM}}^{(1)} = A_{1,1} \text{Im}(\mathbf{E}^* \times \mathbf{H}) \tag{5}$$

where the pre-factor $A_{1,1} = \mu k^4 \text{Im}[\gamma_{\text{elec}}^{(1)*} \gamma_{\text{mag}}^{(1)}]/(12\pi\varepsilon c)$ is determined by the electric and magnetic dipolar polarizabilities, $\gamma_{\text{elec}}^{(1)}$ and $\gamma_{\text{mag}}^{(1)}$, of the particle; **E** and **H** are the incident field vectors in the particle center. It follows from Eq. (5) that an azimuthal force can be generated by azimuthal IPM incidence. For those particles that can support multipolar responses, the force associated with the IPM can be formulated, by a multipole moment expansion technique [27-29] and the angular spectrum theorem, as (see Method for details):

$$\mathbf{F}_{\text{IPM}}^{(N)} = \sum_{l=1}^{N} A_{N,l} \left(k^2 + \frac{\Delta}{2}\right)^{l-1} \text{Im}(\mathbf{E}^* \times \mathbf{H}), \tag{6}$$

where $\Delta$ is the Laplacian operator and $A_{N,l}$ is determined by the particle properties [see Eq. (19)]. The positive integer $N$ denotes the truncation index used in actual calculations, that is, the highest order of the multipole included; the larger the value of $N$, the higher-order components of force are included.

*Equation (6) is the second major theoretical result of this paper,* which generalizes the IPM force to the multipolar regime. For dipolar particles ($N = 1$), it reproduces Eq. (5) that is directly proportional to the IPM. However, the Laplacian $\Delta$ involved in the case of $N > 1$ indicates that the proportionality is invalid in the presence of multipoles. Such nonlinear effects, arising from the higher-order terms, can also contribute to the azimuthal force, because the Laplacian action on the azimuthal IPM will not alter its direction, but just impose a modulation on its magnitude: $\Delta^l(\text{Im}\Pi_\phi \mathbf{e}_\phi) = \mathbf{e}_\phi \tilde{\Delta}^l \text{Im}\Pi_\phi$, where $\tilde{\Delta} \equiv [(\partial_\rho)^2 + (1/\rho)\partial_\rho - 1/\rho^2 + (\partial_z)^2]$. Additionally, one may deduce from Eqs. (18) in Methods that the intensity-gradient force and the radiation pressure have no azimuthal component for the fields shown in Fig. 1b. It is evident from the above considerations that the IPM vortex beam is an ideal platform to isolate the IPM force in the azimuthal direction, even in the multipolar scenario.

To know the multipolar effects excited by the IPM vortex beam, optical forces were calculated by the Lorenz-Mie theory [30], for a gold sphere placed on the focal plane of the field with $\alpha = -45°$ (i.e., the IPM vortex condition). The radius and refractive index of the sphere is set as 0.75 μm and $0.26 + 6.97i$ (the value at $\lambda = 1064$ nm). Figure 2a presents the total azimuthal and radial forces acting on the particle. Two equilibrium positions (I and II) can be identified, where $F_\phi$ is significant and takes opposite signs. It means that the particle tends to rotate clockwise and anticlockwise about the beam axis, for the trapping positions I and II, respectively.

As discussed above, $F_\phi$ is provided by the IPM force. Figure 2a shows the dipolar component of $F_\phi$, calculated with truncation index $N = 1$, together with the distribution of the azimuthal IPM of illumination. It stands out that the dipolar component is proportional to the IPM, as expected from Eq. (5), but it is negligible at the equilibrium positions where the IPM is vanishingly small. However, the proportional relationship is absent for $N > 1$, as shown in Fig. 2c. This can be explained by the high-order terms in Eq. (6), which are nonlinear with respect to the IPM. Thanks to the nonlinear effect, a marked azimuthal force can be produced at each trapping position for $N > 4$. Such a radial trapping, with suppressed low-order IPM components, facilitates our experimental observation of the high-order IPM force.

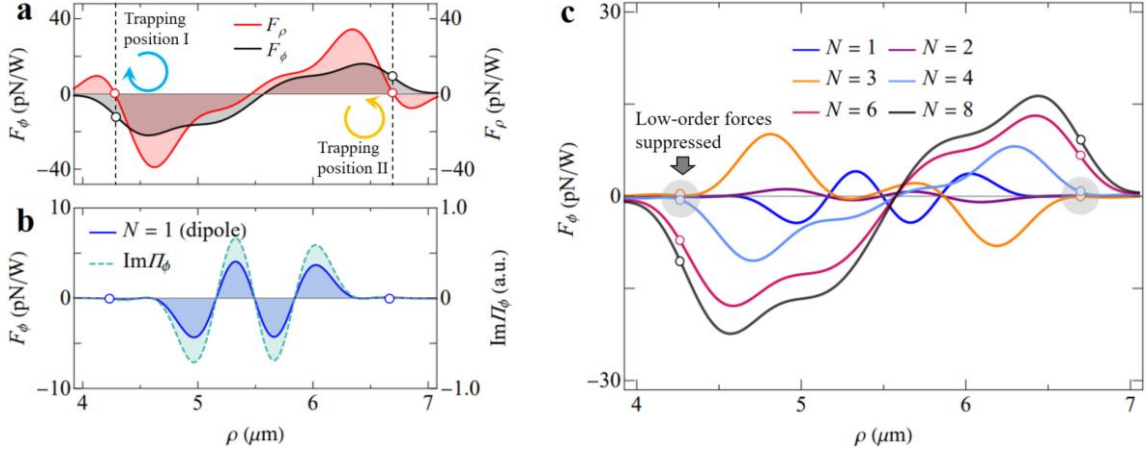

**Figure 2. Numerical results of optical forces for $\alpha = -45°$ (the IPM vortex condition). a,** Radial and azimuthal forces, $F_\rho$ and $F_\phi$, versus the radial displacement of the gold sphere. The truncation index in these calculations is $N = 11$, which is large enough to ensure the convergence of the Mie series. Hollow dots mark the values at radial equilibrium positions, where the radial force vanishes with a negative derivative. **b**, Comparision of the dipolar component of $F_\phi$ with incident IPM. **c**, Azimuthal force for different $N$ showing that low-order contributions are small at the equilibrium positions.

**Experimental observation of the high-order IPM force.** The experimental configuration is shown in Fig. 3a. A spatial light modulator (SLM) and a polarization converter (S-Waveplate) are used to modulate the input field phase and polarization, respectively (see Method for detailed descriptions). The hologram loaded onto the SLM is shown in inset I, which is consistent with that used in our theoretical calculations. Inset II shows the measured intensity distribution at the focal plane, featuring a doubling-ring profile with radius about 5.5 μm. In line with the numerical simulations, we use Au spheres (diameter: 1.5 μm; immersed in water) to probe the optical force. These metallic particles can experience repulsive radiation pressure in the longitudinal direction, which cannot be overcome by the intensity-gradient force and their gravity. In our experiments, the longitudinal repulsive effect was compensated by the surface of the glass cover, so that the motion of the spheres is limited in two dimensions.

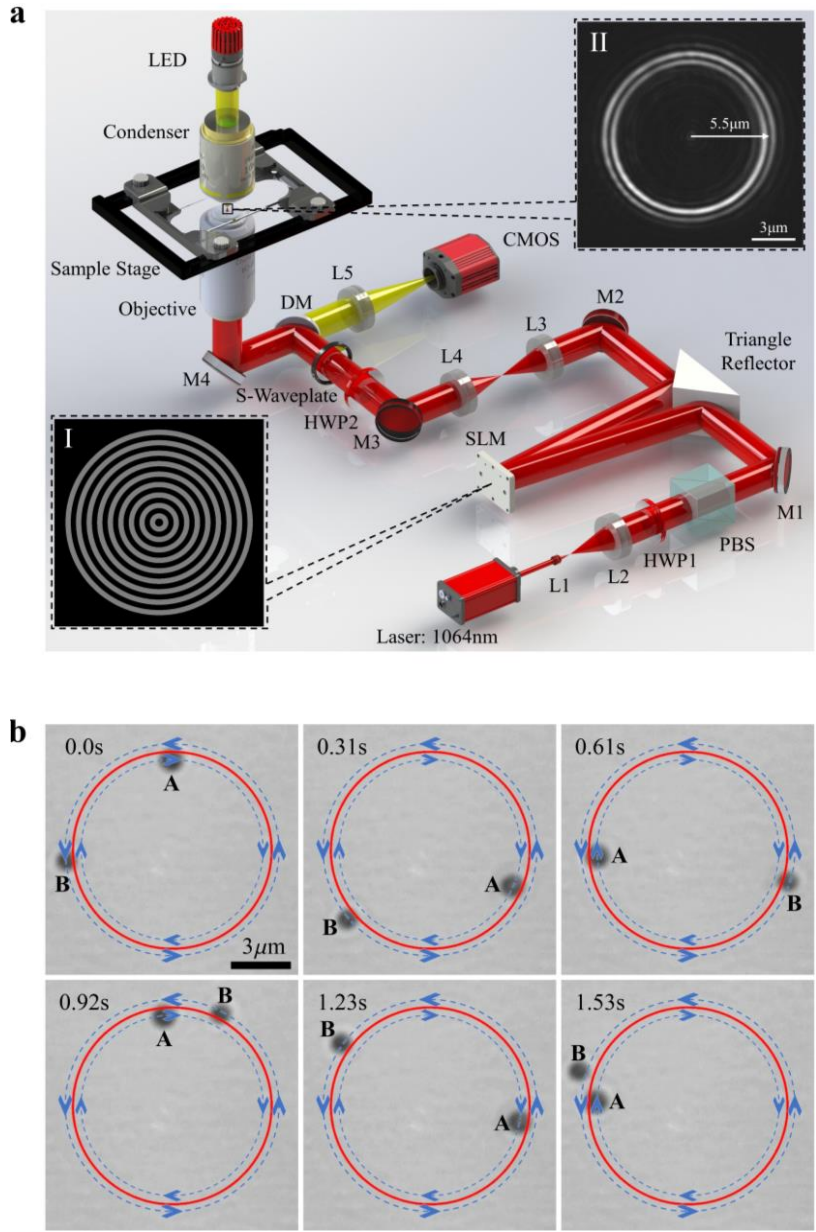

**Figure 3. Particle rotation in the focused IPM vortex beam. a,** Schematics of experimental set-ups. L, Lens; HWP, half-wave plate; PBS, polarizing beam splitter; M, mirror; SLM, spatial light modular; DM, dichroic mirror; CMOS, complementary metal oxide semiconductor camera. The polarization parameter $\alpha$ is controlled by S-Waveplate. Insets I and II show the phase mask profile on the SLM and the measured intensity profile at the focal plane, respectively. **b,** Successive images showing two Au particles trapped and rotated by the IPM vortex beam with polarization parameter $\alpha = -45°$. Red circles with a radius of 5.5 μm illustrate the beam profile. The rotation direction of the spheres is indicated by arrows.

We exemplify in Fig. 3b (Supplementary Video 1) the off-axis dual trapping using two spheres (A and B), which are radially confined, inner and outer respectively, to the red circle that illustrates the focused beam. For both the inner (IRT) and outer radial trapping (ORT), the spheres are shown to revolve about the beam axis, as a manifestation of the high-order IPM force which acts in the azimuthal direction.

However, their rotational behaviors are quite different: the sphere A rotates clockwise, while the sphere B does it anticlockwise. The rotation directions, opposite with respect to each other, agree well with our theoretical results that the azimuthal forces at the two distinct positions of radial equilibrium have opposite signs (Fig. 2c). This rotation ability of the IPM vortex beam is in sharp contrast with that of the conventional optical spanner based on a phase vortex [31-33], in which the particles usually orbit in the same sense.

It was shown previously that metallic microparticles in tightly focused beams with a point focal pattern may generate bubbles due to light-induced heating [34]. We did not observe the bubble formation in the experiments, suggesting that the ring-like focal pattern of the IPM vortex beam alleviates the heating effects. The particle could also undergo a photophoretic force produced by inhomogeneous heating of the particle's surface [35]. Nevertheless, a careful consideration (Supplementary Note 2) shows that the photophoretic force is small in our case and its azimuthal component is opposite to the rotation direction of the particle, for both the IRT and ORT. For these reasons, the particle rotation can be attributed exclusively to the azimuthal IPM force.

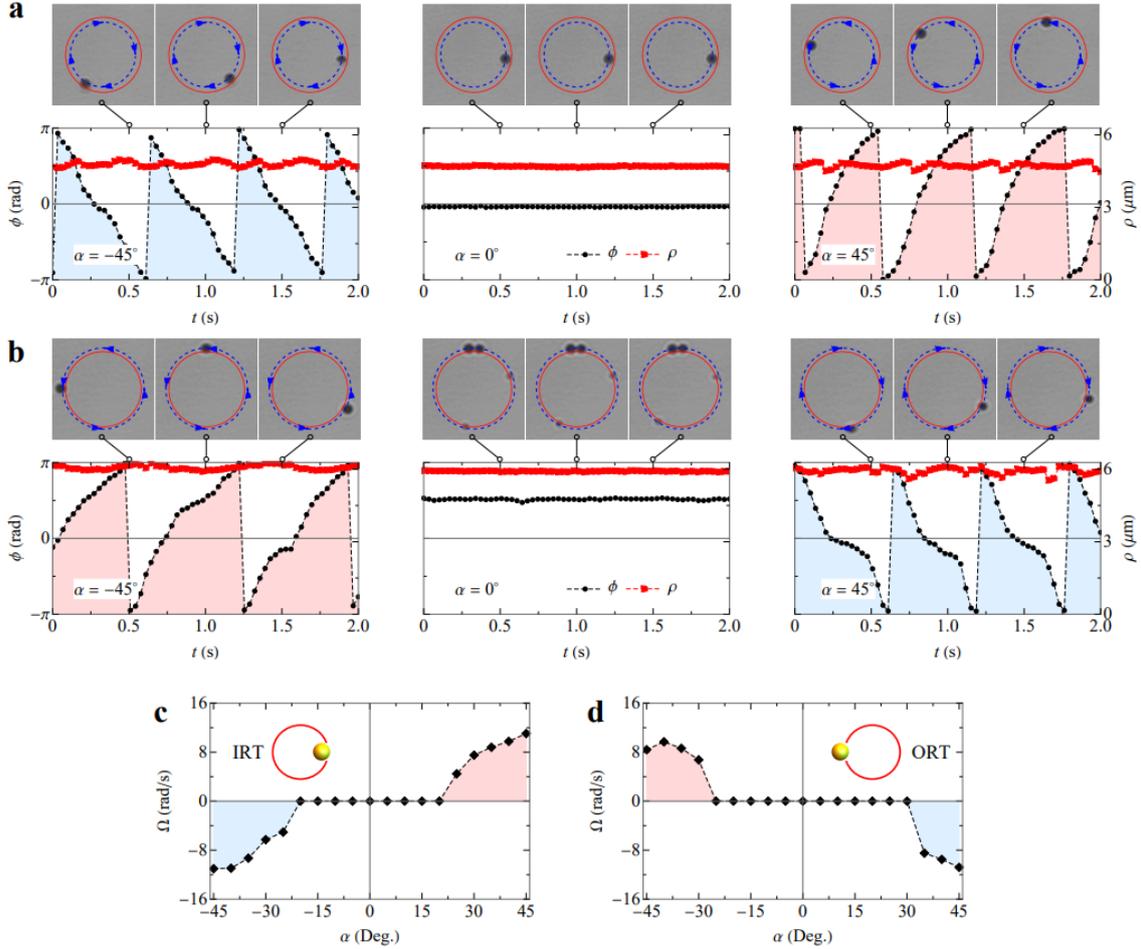

**Figure 4. Dependence of the particle rotation on the polarization. a,b,** Real-time positions of a single sphere for the IRT (**a**) and ORT (**b**) for different values of the polarization parameter $\alpha$. Insets show experimental snapshots of the rotating sphere at selected time points. **c,d,** Measured angular speed of the sphere versus $\alpha$ for the IRT (**c**) and ORT (**d**).

To experimentally examine the polarization-dependent property of the IPM force, the particle's trajectories were recorded at varying $\alpha$. Note that, here, a single sphere is used in order to avoid possible interparticle interactions. Figure 4a shows the results for the IRT (Supplementary Video 2). The clockwise motion occurring at $\alpha = -45°$ is halted when $\alpha$ is switched to 0, and a rotation of reversed direction is driven at $\alpha = 45°$. The sensitivity of the rotation to the field polarization is also remarkable for the ORT (Fig. 4b, Supplementary Video 3). However, the ORT yields a rotation direction which is always opposite to that for the IRT, as it should be. Figures 4c and d show the experimentally measured averaged angular speed $\Omega$ of the sphere as a function of $\alpha$, for the IRT and ORT, respectively. The speed profile in either case is almost symmetric, with a sinusoidal-like region around $\alpha = \pm 45°$. This is a typical signature of the IPM force, as can be seen by the factor $\sin(2\alpha)$ in Eq. (4). However, we did not observe a significant rotation when $|\alpha|$ or $|\sin(2\alpha)|$ is small, although the azimuthal IPM of illumination can be nonzero in these conditions. We ascribe the stationary phenomena to the resistance arising from the glass cover. Namely, since the sphere is trapped against the surface of the glass cover, it should experience a frictional force that could surpass the IPM force, especially when $|\sin(2\alpha)|$ is small.

**Discussions**

We have demonstrated, both numerically and experimentally, the direct observation of the IPM force, by utilizing the IPM vorticity in a structured light beam and the high-order Mie responses of gold microparticles. It indicates that the IPM force, a long time overlooked mechanical effect of light [4,16,36], is actually ubiquitous, being exerted on a large variety of Mie particles [37,38]. This force is shown to be remarkable, as the dynamics of the particles can be effectively controlled by this force alone. It functions in the beam as a peculiar optical spanner that sets the particles into bidirectional rotation, which endows itself with novel characteristics for applications in practical optical manipulation.

Moreover, our theoretical model describes the multipolar contributions to the optical force derived from various field quantities of interest. It tells us that the optical force on a generic Mie particle can be classified into four fundamental types, which are associated with the intensity gradient, canonical momentum, time-averaged momentum and imaginary Poynting momentum, respectively. This is akin to the landscape in the dipolar approximation [9-13]. However, one should draw special attention to those high-order terms, which are nonlinear with respect to their corresponding field quantities. Such nonlinear effects may open new opportunities for tailoring light-matter interactions, and should empower advanced applications in plasmonics and in the emerging field of Mie-tronics [39].

As a closing remark, we stress that the IPM force can arise from various multipolar interplay processes. In the dipole limit, this interaction is simply the coupling between the electric and magnetic dipoles [9-14]; for a generic Mie particle, the interactions also include the electric-electric coupling and magnetic-magnetic coupling. From Eqs. (18)-(20) in Methods, we see that $\mathbf{F}_{\text{IPM}}^{\text{e}(N)}$ (or $\mathbf{F}_{\text{IPM}}^{\text{m}(N)}$) represents the purely electric (or magnetic) IPM force resulting from the interaction of electric (or magnetic) multipoles with adjacent orders. One may use $\mathbf{F}_{\text{IPM}}^{\text{e}(N)} - \mathbf{F}_{\text{IPM}}^{\text{e}(N-1)}$ (or $\mathbf{F}_{\text{IPM}}^{\text{m}(N)} - \mathbf{F}_{\text{IPM}}^{\text{m}(N-1)}$) to calculate the component due to the interplay of the electric (or magnetic) $2^N$-pole with $2^{N-1}$-pole. For example, $\mathbf{F}_{\text{IPM}}^{\text{e}(2)} - \mathbf{F}_{\text{IPM}}^{\text{e}(1)}$ evaluates the electric quadrupole-dipole IPM force. On the other hand, $\mathbf{F}_{\text{IPM}}^{\text{x}(N)}$ is the hybrid magnetoelectric IPM force derived from the interference between electric and magnetic multipoles of the same order, and $\mathbf{F}_{\text{IPM}}^{\text{x}(N)} - \mathbf{F}_{\text{IPM}}^{\text{x}(N-1)}$ gives the hybrid $2^N$-pole IPM force.

**Methods**

**Phase modulation on the input field.** The amplitude function of the CV input field can be expressed as

$$\mathcal{A}_0(\vartheta) = E_0 \mathcal{P}_{\text{Mod}}(\vartheta) e^{-\beta_0^2 (\sin\vartheta/\sin\Theta)^2} J_1\left(2\beta_0 \frac{\sin\vartheta}{\sin\Theta}\right), \qquad (7)$$

where $E_0$ is the peak field amplitude and a small parameter $\beta_0$ is typically used ($\beta_0 = 1.5$ in our case) so that the input field can fill the aperture. The radial modulation function, $\mathcal{P}_{\text{Mod}}(\vartheta)$, is used to shape the focused field. The case $\mathcal{P}_{\text{Mod}}(\vartheta) = 1$ corresponds to a normal CV input field [23], tight focusing of which will cause the obtained IPM vortex to be concentrated in an immediate vicinity of the beam axis (see Supplementary Note 3 for details). To localize the field away from the axis, we have resorted to perfect vortex generation technique, in which an additional radially varying modulation factor is introduced, to create an annular intensity profile [40]. For our application, we use

$$\mathcal{P}_{\text{Mod}}(\vartheta) = \text{sign}\left[J_0(k_1 \rho_0 \sin\vartheta)\right]. \qquad (8)$$

This phase modulation will split the input field into a diverging spherical wave and a converging spherical wave, thereby their interfering results in an annular pattern in the focal plane. The focused field obtained by this modulation factor will be of an annular pattern of radius $\sim \rho_0$. In our work, $\rho_0$ is set to be 5.5 μm.

**High-order theory of IPM force, intensity-gradient force and radiation pressure.** Our theory is developed based on the optical force model established by Lin et al. [27-29,41,42]. In this model, the optical force on a generic sphere is described by a set of incident field moments:

$$\begin{aligned}
\mathbf{S}_{\text{em}}^{(n)} &= \left[(\nabla^{(n-1)} \mathbf{E}^*) \overset{(n-1)}{:} (\nabla^{(n-1)} \mathbf{H})\right] \overset{(2)}{:} \boldsymbol{\epsilon}^{(3)}, \\
D_{\text{e}}^{(n)} &= (\nabla^{(n-1)} \mathbf{E}^*) \overset{(n)}{:} (\nabla^{(n-1)} \mathbf{E}), \\
D_{\text{m}}^{(n)} &= (\nabla^{(n-1)} \mathbf{H}^*) \overset{(n)}{:} (\nabla^{(n-1)} \mathbf{H}), \\
\mathbf{S}_{\text{e}}^{(n)} &= \left[(\nabla^{(n-1)} \mathbf{E}^*) \overset{(n-1)}{:} (\nabla^{(n-1)} \mathbf{E})\right] \overset{(2)}{:} \boldsymbol{\epsilon}^{(3)}, \\
\mathbf{S}_{\text{m}}^{(n)} &= \left[(\nabla^{(n-1)} \mathbf{H}^*) \overset{(n-1)}{:} (\nabla^{(n-1)} \mathbf{H})\right] \overset{(2)}{:} \boldsymbol{\epsilon}^{(3)},
\end{aligned} \qquad (9)$$

with $\boldsymbol{\epsilon}^{(3)}$ the Levi-Civita tensor and the integer $1 \leq n \leq N$ denoting the order of the moments. Although the model involves complicated formulae, it is not difficult to observe from Ref. [28] or [29] that the force can be arranged into a series over these moments:

$$\mathbf{F}^{(N)} = \sum_{l=1}^{N} \left\{ \begin{array}{l} A_{N,l} \mathrm{Im} \mathbf{S}_{\mathrm{em}}^{(l)} + B_{N,l} \nabla D_{\mathrm{e}}^{(l)} + C_{N,l} \nabla D_{\mathrm{m}}^{(l)} + \\ D_{N,l} \mathrm{Re} \mathbf{S}_{\mathrm{em}}^{(l)} + E_{N,l} \mathbf{P}_{\mathrm{e}}^{(l)} + F_{N,l} \mathbf{P}_{\mathrm{m}}^{(l)} \end{array} \right\}. \tag{10}$$

where the truncation index $N$ denotes the highest order of multipole included, and we have introduced the moments:

$$\mathbf{P}_{\mathrm{e}}^{(n)} = \omega \, \mathrm{Re} \mathbf{S}_{\mathrm{em}}^{(n)} + i \nabla \times \mathbf{S}_{\mathrm{e}}^{(n)}/2, \quad \mathbf{P}_{\mathrm{m}}^{(n)} = \omega \, \mathrm{Re} \mathbf{S}_{\mathrm{em}}^{(n)}/c^2 + i \nabla \times \mathbf{S}_{\mathrm{m}}^{(n)}/2. \tag{11}$$

The coefficients $A_{N,l} \sim F_{N,l}$ are determined by both $N$ and the polarizabilities of the particle. For $N = 1$, Eq. (10) gives the well-known dipolar optical force [13], which can be decomposed into the IPM force, intensity-gradient force and radiation pressure. In the following we will generalize these concepts to the case of $N > 1$.

To proceed, we turn to $\mathbf{k}$-space, in which the incident fields $\mathbf{E}$ and $\mathbf{H}$ are expressed by plane-wave components [16,43,44]:

$$\mathbf{E} = \int_{-\infty}^{+\infty} \mathbf{f}(\mathbf{u}) e^{i\mathbf{k}\cdot\mathbf{x}} \mathrm{d}^2 K, \quad \mathbf{H} = \int_{-\infty}^{+\infty} \mathbf{h}(\mathbf{u}) e^{i\mathbf{k}\cdot\mathbf{x}} \mathrm{d}^2 K, \tag{12}$$

where $\mathbf{k} = (k_x, k_y, k_z)$, $|\mathbf{k}| = k$, $\mathbf{u} = (k_x, k_y)$, and $\mathrm{d}^2 K = \mathrm{d}k_x \mathrm{d}k_y$. Using this representation, we find

$$\mathbf{S}_{\mathrm{em}}^{(n)} = \iint (\mathbf{k}_1^* \cdot \mathbf{k}_2)^{n-1} [\mathbf{f}^*(\mathbf{u}_1) \times \mathbf{h}(\mathbf{u}_2)] e^{-i\mathbf{k}_{12}\cdot\mathbf{x}} \mathrm{d}^2 K_1 \mathrm{d}^2 K_2, \tag{13}$$

with $\mathbf{k}_{12} \equiv \mathbf{k}_1^* - \mathbf{k}_2$. The key step is to note the equality: $(\mathbf{k}_1^* \cdot \mathbf{k}_2) = k^2 - (\mathbf{k}_{12})^2/2$, which yields

$$(\mathbf{k}_1^* \cdot \mathbf{k}_2)^{n-1} = \left[ k^2 - \frac{(\mathbf{k}_{12})^2}{2} \right]^{n-1} = \sum_{j=0}^{n-1} C_{n,j}(k) (\mathbf{k}_{12})^{2j}, \tag{14}$$

where

$$C_{n,j}(k) = \left( -\frac{1}{2} \right)^j \frac{(n-1)!}{j!(n-1-j)!} k^{2n-2j-2}.$$

Upon substitution, we have

$$\mathbf{S}_{\mathrm{em}}^{(n)} = \sum_{j=0}^{n-1} C_{n,j}(k) \iint (\mathbf{k}_{12})^{2j} [\mathbf{f}^*(\mathbf{u}_1) \times \mathbf{h}(\mathbf{u}_2)] e^{-i\mathbf{k}_{12}\cdot\mathbf{x}} \mathrm{d}^2 K_1 \mathrm{d}^2 K_2, \tag{15}$$

We observe that the role of $(\mathbf{k}_{12})^{2j}$ in the integrand is equivalent to the negative Laplacian $-\Delta = -\nabla^2$, which is operated in the $\mathbf{x}$-space. Eq. (15) can thus be rewritten as

$$\mathbf{S}_{\mathrm{em}}^{(n)} = \sum_{j=0}^{n-1} (-\Delta)^j C_{n,j}(k) \iint [\mathbf{f}^*(\mathbf{u}_1) \times \mathbf{h}(\mathbf{u}_2)] e^{-i\mathbf{k}_{12}\cdot\mathbf{x}} \mathrm{d}^2 K_1 \mathrm{d}^2 K_2$$

$$= \sum_{j=0}^{n-1} (-\Delta)^j C_{n,j}(k) \mathbf{S}_{\mathrm{em}}^{(1)} \tag{16}$$

$$= \left( k^2 + \frac{\Delta}{2} \right)^{n-1} \mathbf{S}_{\mathrm{em}}^{(1)}.$$

In a development paralleling the steps from Eqs. (12) to (16), we obtain the results for the other field moments in Eq. (10):

$$D_{\mathrm{e}}^{(n)} = \left(k^2 + \frac{\Delta}{2}\right)^{n-1} D_{\mathrm{e}}^{(1)}, \quad D_{\mathrm{m}}^{(n)} = \left(k^2 + \frac{\Delta}{2}\right)^{n-1} D_{\mathrm{m}}^{(1)},$$
$$\mathbf{P}_{\mathrm{e}}^{(n)} = \left(k^2 + \frac{\Delta}{2}\right)^{n-1} \mathbf{P}_{\mathrm{e}}^{(1)}, \quad \mathbf{P}_{\mathrm{m}}^{(n)} = \left(k^2 + \frac{\Delta}{2}\right)^{n-1} \mathbf{P}_{\mathrm{m}}^{(1)}. \tag{17}$$

Equations (16) and (17) link the field moments of arbitrary orders to their lowest order counterparts, whose physical meaning is quite straightforward: $D_{\mathrm{e}}^{(1)} = |\mathbf{E}|^2$ (electric intensity); $D_{\mathrm{m}}^{(1)} = |\mathbf{H}|^2$ (magnetic intensity); $\mathbf{P}_{\mathrm{e}}^{(1)} = \mathrm{Im}[\mathbf{E}^* \cdot (\nabla) \mathbf{E}]$ (electric orbital momentum); $\mathbf{P}_{\mathrm{m}}^{(1)} = \mathrm{Im}[\mathbf{H}^* \cdot (\nabla) \mathbf{H}]$ (magnetic orbital momentum); $\mathbf{S}_{\mathrm{em}}^{(1)} = \mathbf{E}^* \times \mathbf{H}$ (complex Poynting vector). Substituting these results into Eq. (10), and according to field-related characteristics, one may categorize the force into four parts: $\mathbf{F}^{(N)} = \mathbf{F}_{\mathrm{IPM}}^{(N)} + \mathbf{F}_{\mathrm{G}}^{(N)} + \mathbf{F}_{\mathrm{RP}}^{(N)} + \mathbf{F}_{\mathrm{CRP}}^{(N)}$, where

$$\mathbf{F}_{\mathrm{IPM}}^{(N)} = \sum_{l=1}^{N} A_{N,l} \left(k^2 + \frac{\Delta}{2}\right)^{l-1} \mathrm{Im} \mathbf{S}_{\mathrm{em}}^{(1)},$$
$$\mathbf{F}_{\mathrm{G}}^{(N)} = \sum_{l=1}^{N} \left(k^2 + \frac{\Delta}{2}\right)^{l-1} [B_{N,l} \nabla D_{\mathrm{e}}^{(1)} + C_{N,l} \nabla D_{\mathrm{m}}^{(1)}],$$
$$\mathbf{F}_{\mathrm{RP}}^{(N)} = \sum_{l=1}^{N} D_{N,l} \left(k^2 + \frac{\Delta}{2}\right)^{l-1} \mathrm{Re}[\mathbf{S}_{\mathrm{em}}^{(1)}], \tag{18}$$
$$\mathbf{F}_{\mathrm{CRP}}^{(N)} = \sum_{l=1}^{N} \left(k^2 + \frac{\Delta}{2}\right)^{l-1} [E_{N,l} \mathbf{P}_{\mathrm{e}}^{(1)} + F_{N,l} \mathbf{P}_{\mathrm{m}}^{(1)}].$$

We call $\mathbf{F}_{\mathrm{IPM}}^{(N)}$ the *generalized IPM force* as it originates from the IPM of illumination. The second and third rows can be identified as the *generalized intensity-gradient force* and *generalized radiation pressure*, induced by the optical intensity inhomogeneity and momentum, respectively. But it is worth noting that the electric and magnetic fields contribute differently to these two forces. The last term, $\mathbf{F}_{\mathrm{CRP}}^{(N)}$, is also a type of radiation pressure, as the orbital momentum represents the canonical part of the optical momentum [9]. For this reason, we coin it the *generalized canonical radiation pressure*. We remark that the above generalization is self-consistent, because for $N = 1$, all the terms in Eq. (18) reduce to their dipolar counterparts [13].

So far we have not dealt with the coefficients $A_{N,l} \sim F_{N,l}$. These coefficients can be worked out with proper index substitution for Eqs. (11)-(22) in Ref. [29] (see Supplementary Note 4 for details). As the main goal of this paper is the IPM force, we present the result for $A_{N,l} = A_{N,l}^{\mathrm{e}} + A_{N,l}^{\mathrm{m}} + A_{N,l}^{\mathrm{x}}$:

$$A_{N,l}^{\text{e}} = \frac{\mu\omega}{4\pi\varepsilon}\left\{\sum_{j=0}^{\left\lfloor\frac{N-l-1}{2}\right\rfloor} \Lambda_{l+2j,j}\operatorname{Im}[\gamma_{\text{elec}}^{(l+2j)}\gamma_{\text{elec}}^{(l+2j+1)*}] + \sum_{j=0}^{\left\lfloor\frac{N-l-2}{2}\right\rfloor}\Omega_{l+2j+1,j}\operatorname{Im}[\gamma_{\text{elec}}^{(l+2j+1)}\gamma_{\text{elec}}^{(l+2j+2)*}]\right\},$$

$$A_{N,l}^{\text{m}} = -\frac{\mu\omega}{4\pi\varepsilon c^4}\left\{\sum_{j=0}^{\left\lfloor\frac{N-l-1}{2}\right\rfloor} \Lambda_{l+2j,j}\operatorname{Im}[\gamma_{\text{mag}}^{(l+2j)}\gamma_{\text{mag}}^{(l+2j+1)*}] + \sum_{j=0}^{\left\lfloor\frac{N-l-2}{2}\right\rfloor}\Omega_{l+2j+1,j}\operatorname{Im}[\gamma_{\text{mag}}^{(l+2j+1)}\gamma_{\text{mag}}^{(l+2j+2)*}]\right\},$$

$$A_{N,l}^{\text{x}} = \frac{\mu}{4\pi\varepsilon c}\left\{\sum_{j=0}^{\left\lfloor\frac{N-l}{2}\right\rfloor} M_{l+2j,j}\operatorname{Im}[\gamma_{\text{elec}}^{(l+2j)*}\gamma_{\text{mag}}^{(l+2j)}] - \frac{lk^2}{(l+1)^2}\sum_{j=0}^{\left\lfloor\frac{N-l-1}{2}\right\rfloor} M_{l+2j+1,j}\operatorname{Im}[\gamma_{\text{elec}}^{(l+2j+1)*}\gamma_{\text{mag}}^{(l+2j+1)}]\right.$$

$$\left. -\frac{l(l+1)k^4}{(l+2)^2}\sum_{j=0}^{\left\lfloor\frac{N-l-2}{2}\right\rfloor} M_{l+2j+2,j}\operatorname{Im}[\gamma_{\text{elec}}^{(l+2j+2)*}\gamma_{\text{mag}}^{(l+2j+2)}]\right\},\quad (19)$$

where $\gamma_{\text{elec}}^{(n)}$ and $\gamma_{\text{mag}}^{(n)}$ are the electric and magnetic $2^n$-polar polarizabilities, and the related parameters ($\Lambda_{n,m}$, $\Omega_{n,m}$, and $M_{n,m}$) are given by Eq. (S4.12) in Supplementary Information. To gain an insight into the physics underneath the IPM force, we have decomposed $A_{N,l}$ into three parts that correspond to different force contributions:

$$\mathbf{F}_{\text{IPM}}^{(N)} = \mathbf{F}_{\text{IPM}}^{\text{e}(N)} + \mathbf{F}_{\text{IPM}}^{\text{m}(N)} + \mathbf{F}_{\text{IPM}}^{\text{x}(N)} = \sum_{l=1}^{N} A_{N,l}^{\text{e}}\operatorname{Im}\mathbf{S}_{\text{em}}^{(l)} + \sum_{l=1}^{N} A_{N,l}^{\text{m}}\operatorname{Im}\mathbf{S}_{\text{em}}^{(l)} + \sum_{l=1}^{N} A_{N,l}^{\text{x}}\operatorname{Im}\mathbf{S}_{\text{em}}^{(l)}. \quad (20)$$

As an illustration, we consider the case shown in Fig. 2c and make a decomposition of the azimuthal IPM force into the three parts. The results are shown in Supplementary Fig. S3.

**Experimental setup.** Our experiments were performed based on home-built holographic optical tweezers (Fig. 3a). A linearly polarized beam ($\lambda$ = 1064 nm) was expanded and collimated by a telescope consisting of L1 and L2. After passing through a half-wave plate (HWP1) and a polarizing beam splitter (PBS), the input beam becomes horizontally polarized. To miniaturize the size of the setup, a specially designed 96° triangle reflector is employed to reflect the input beam onto an SLM (Pluto-HED6010-NIR-049-C, Holoeye Photonics AG Inc., Germany, 1920×1080 pixels, pixel pitch: 8.0 μm, frame rate: 60 Hz). The modulated beam was relayed into the back aperture of the objective (100×, NA 1.4, Oil-immersion, CFI Plan Apo, Nikon Inc., Japan) by a 4*f* system consisting of L3 and L4. After L4, a half-wave plate (HWP2) and an S-Waveplate (RPC-1064-08-334, Workshop of Photonics Inc., Lithuania) were used to convert the linearly polarized beam into a vector beam.

It is known that the S-Waveplate can be considered as a half-wave plate with space variant direction of fast-axis. To control the polarization direction, i.e., the polarization angle $\alpha$, of the generated vector beam with respect to the radial direction, the HWP2 was arranged into a stepper motor rotation mount (K10CR1/M, Thorlabs Inc., USA) to rotate the polarization direction of the linearly polarized beam incident onto the S-Waveplate. The generated beam can be considered as a linearly polarized vector Bessel-Gauss beam and then focused by the objective. The objective was also employed for the sample imaging, and a CMOS camera (Point Grey GS3-U3-41C6M-C, FLIR

System Inc., USA, 2048×2048 pixels, pixel pitch: 5.5 μm, frame rate: 90 fps) was used to monitor and record the manipulation process.

## Author contributions

X.X., S.Y., C.-W.Q. and B.Y. conceived the original concept and initiated the work. Y.Z. and Y.Z. performed the experiments, under the supervision of B.Y.. X.X., S.Y. and M.N.-V. developed the high-order theory of optical force. M.L. and S.Y. conducted the numerical simulations, with the assistance of Y.Z. and B.L.. X.X., S.Y., M.N.-V and C.-W.Q. wrote the paper, and all authors reviewed the manuscript.

## Acknowledgements

This work is supported by the National Natural Science Foundation of China (Nos. 11974417, 62135005, 11904395, 12127805, and 11804119) and the Key Research Program of Frontier Sciences, CAS (No. ZDBS-LY-JSC035). M.N.-V acknowledges Ministerio de Ciencia e Innovación of Spain, research grant PGC2018-095777-B-C21.